# S-BLE: A Participatory BLE Sensory Data Set Recorded from Real-World Bus Travel Events

Jonathan Lam, Roberto Manduchi
University of California, Santa Cruz

**ABSTRACT**
This contribution describes S-BLE, a data set created for supporting the design of robust and reliable Be-In-Be-Out systems in public transit. S-BLE was recorded by the smartphones of 28 participants during their daily transit routines in a university campus setting. 20 shuttle bus vehicles in the campus fleet were equipped with two Bluetooth low energy (BLE) beacons each. RSSI data from these beacons was recorded during regular rides, along with odometry information (from GPS) and data from the smartphone's inertial sensors. The article describes the system used for data collection and presents some statistics of interest for the recorded data.

**INTRODUCTION**
Be-In-Be-Out (BIBO [1]) refers to mechanisms by which passengers can interact with a transit system without the need for a specific action (such as handling paper tickets, tagging or presenting a QR code to a reader using a card or a smartphone, or inputting information in a smartphone app). BIBO systems are often studied in the context of *implicit ticketing* [1] (or *frictionless transit* [2]) which can be attractive to passenger because of its convenience and because it may reduce the risk of paying an excessive fare (e.g., when passengers forget to tag out at the end of a trip when the agency uses a tag-in, tag-out systems.) Efficient payment systems including implicit ticketing may play a critical role in enhancing the profitability of transit services [3]. BIBO interaction can also facilitate other data collection tasks, including occupancy/crowdedness tracking or unplanned disturbance management [4].

At its core, a BIBO system is tasked with detecting when a passenger steps inside or outside of a transit vehicle. Methods to accomplish this without additional infrastructure have been proposed in the literature (e.g., by matching the time series of sensor data collected by the passenger's smartphone with the same data recorded by the transit vehicle [5] [6] [7] [8]), but their robustness and reliability has not been tested in real-world situations. A growing body of work has considered adding a "light" infrastructure layer based on Bluetooth Low-Energy (BLE) signaling. BLE beacons are small, relatively inexpensive, and use very little energy. Since their transmission radius is limited, an approximate measure of the user's distance to the beacon can be obtained from the strength of the signal received from the beacon (received signal strength indicator, or RSSI) [9]. All modern smartphones are already equipped with BLE transceivers, and thus can be used to discover and read data from BLE beacons. Signal received from a beacon with known ID, possibly combined with GPS or inertial data (e.g., to measure the passenger's speed), can be used to verify whether or not a user is inside a transit vehicle.

Representative related research work includes early investigation by Narzt et al. [1], who experimented with a BIBO system in which users themselves advertise a beacon ID by means of a BLE beacon device or a BLE-enabled smartphone, while one or more BLE receivers were installed in a bus vehicle and communicated with a remote server. The authors reported RSSI measured with different configurations (number of receivers, beacon location, antenna orientation), with measurements taken in an empty bus as well as in a full bus. Later work considered a different configuration, with BLE beacons installed in the vehicle, and the user's own smartphone receiving data from the beacons and connecting to a remote back-end. Ferrera et al. [10] [11] presented a pilot test with the ANDA system, deployed in two bus lines, a light rail line, and a rail line in the metropolitan area of Porto (Portugal). BLE beacons were installed in the bus vehicles (one beacon per vehicle), while for the light rail and rail they were installed at a number of stations. The system was designed such that the information broadcasted by the in-bus beacons would change dynamically during the trip. An Android mobile app was developed that tracked passengers through their journeys and implemented a post-billing system. Power measurements were reported from specific experiments, along with feedback from a cohort of 90 participants who volunteered to test the system. A similar approach was taken by Sarkar et al. [12], who focused on the security, privacy, and energy efficiency aspects of the system. Recent work by Servizi et al. [13] [14] described a system that uses deep learning for unsupervised, autovalidated BIBO classification of bus passengers, using data from GPS and BLE signals received from bus-carried BLE beacons (one beacon per bus vehicle). Results from a large-scale experiment with 134 bus users in a campus setting were reported. Beep4me [15] is a BIBO system using three BLE beacons in selected bus vehicles, with an additional geofencing area defined at bus stops. Mirzaei and Manduchi [16] showed that not only the presence of a passenger in a bus, but also their approximate location in the vehicle, can be measured from the RSSI measured by the passenger's smartphone from two BLE beacons, located at opposite ends of the bus.

This prior work showed the feasibility of BLE beacons as an infrastructure enabling BIBO interactions, but also brought to light the challenges with such a system when deployed in the "real world". Signal from a beacon placed inside a vehicle can almost certainly be detected by a smartphone carried by a passenger in that vehicle, provided that the beacon was set to a high enough transmission power. However, one must also consider cases in which a beacon is detected even when the user did not actually enter the vehicle (e.g., from a vehicle passing by while the user is waiting at the bus stop). Likewise, the phone of a passenger riding a vehicle may detect beacons from a nearby vehicle. These situations may trigger "false positives", possibly resulting in incorrect fare collection. This may be a reason why the few BLE-based ticketing systems that have been deployed by transit agencies either assume passengers walking through a gate (e.g., tagless gates in South Korea subway lines [17]) or require passengers to check in (via NFC, as in the hybrid Anda system by the fare collection company TIP in Porto, Portugal [18], or by confirming presence in the vehicle on their app, as in the Hamburg, Germany, transit agency HVV [19]).

This contribution describes a new data set (Shuttle-BLE, or S-BLE for short) that was collected with the purpose of supporting the design of reliable and robust BLE-based BIBO systems. S-BLE, which is openly available at **https://doi.org/10.7291/D1BQ2Q**, contains multiple sources of sensory data recorded over a period of ten months (from May of 2023 to February of 2024) by the smartphones of 28 students of the UC Santa Cruz (UCSC) campus who participated in the data collection. These students made regular use of the UCSC campus shuttle bus system, whose 20 vehicles were equipped with two BLE beacons each. Participants installed an iOS app (SBLE-iOS) that automatically started recording as soon as a beacon contained in any one of these shuttle busses was detected. In addition, the app produced prompts asking participants to confirm their presence in the bus, their approximate location in the vehicle, and whether they had left the vehicle. The data collected included the BLE beacon IDs, the received RSSI, the user's location and velocity (via GPS), and other data from the smartphone's inertial sensors. Data was recorded while participants were attending to their daily routine, and thus can be considered representative of "real world" situations. We believe that this data set can be useful for designers of BIBO systems, and in particular for systems based on machine learning [13] [14], which require large amounts of data for training and validation.

In the next sections, we describe the S-BLE system used for data collection, then present some statistics of interest of the recorded data. The limitations of this data set are then discussed in the concluding section.

**METHODS**

**System**
The S-BLE system for data collection has five components:
- BLE beacons placed in 20 UC Santa Cruz campus shuttle busses;
- A back-end web application, hosted by the IBM Cloud Engine, which uses the Flask framework to handle website access and Application Program Interfaces (APIs) for the mobile app, and to track user accounts.
- A NoSQL document database, hosted by IBM Cloudant, which stores and allows for easy retrieval of all collected data.

- A website, which allows users to register an account, login, search, and download their collected data, and view the leaderboard (which tracks and ranks the total number of data collection sessions).
- An iOS smartphone app (SBLE-iOS), which collects data when a BLE beacon on any campus shuttle bus is within range, accepts user input, and displays the leaderboard.

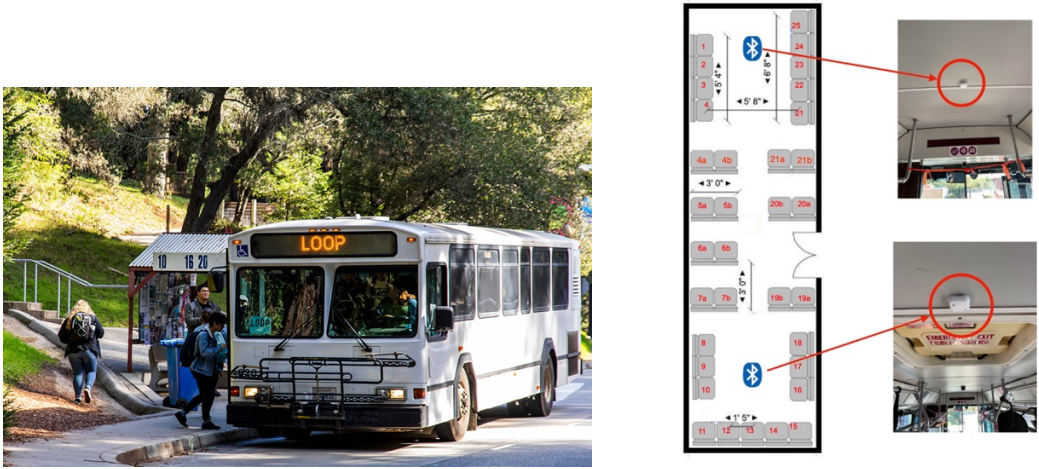

**Figure 1. Left: A UC Santa Cruz campus shuttle bus (photo by Stephen Louis Marino). Right: The map of the passenger area of shuttle bus. The location of the BLE beacons is shown, along with sample pictures of the installed beacons.**

*BLE Beacon Placement*

We placed two BLE beacons (Kontakt Anchor Beacon 2) in each one of the 20 shuttle busses operated by the UCSC Transportation & Parking Services (TAPS; https://taps.ucsc.edu). See Figure 1. These beacons are small (29 mm X 29 mm X 18.5 mm) and have a replaceable battery. They broadcast their ID, along with battery level and some other system data. The bus vehicles in the UCSC fleet are approximately 9 meters long and 2.5 meters wide. They have two entrance doors, one next to the driver (not visible in the passenger area map shown in Figure 1), and one in the middle. The vehicles have enough seating for 34 persons; however, seating regularly fills up during rush hours (e.g., class changing times), and several additional standing passengers are often present in the bus. These shuttle busses operate on a route that forms a 7.1 Km (4.4 miles) loop around the UCSC campus (see Figure ). Half of the busses in the fleet drive through the route clockwise (West to East), and half counterclockwise (East to West). Rides are free for all, and there is no ticketing system. The UCSC campus is located in a rural area, with buildings scattered throughout a 2,000 acres (809 hectares) extension. The North portion of campus is characterized by a dense redwood tree forest.

Following [16], we decided to place two beacons in each bus vehicle, attached to the vehicle's ceiling in the front and in the rear area (see Figure 1, right). All beacons placed on the campus shuttle busses were assigned the same distinctive UUID (Universally Unique Identifier). The SBLE-iOS app, described in the next section, is designed to only range for this pre-set UUID. Besides UUID, BLE beacons have two additional characteristic identifiers: *major* and *minor*. Major IDs are normally used to identify a group of beacons. In our case, the two beacons within each vehicle share the same unique major, and are characterized by different minors. All BLE beacons were set to an advertisement period of 350 ms, with transmission power of level of $-16$ dBm (25 $\mu$W). According to the manufacturer (Kontakt), this power level ensures an approximate range of 10 meters. The actual range, however, depends on the environment in which beacons are deployed. In our case, reflection off the vehicle's walls, ceiling and floor may contribute to increasing the range, while the presence of obstacles (other passengers, seats) may reduce the received power and thus the effective range. The manufacturer states that, at an advertisement period of 1 second and power level of −12 dBm (63 $\mu$W), a beacon's battery life is approximately 8 years. Based on our chosen advertisement period (350 ms) and power (−16 dBm), one may expect a similar battery life for our system.

*SBLE-iOS App*
SBLE-iOS is an iOS app that was designed specifically for this data collection. It records multiple sources of data when in the proximity of a BLE beacon with pre-set UUID, and accepts limited input from the user. We chose to develop the app for iOS considering that the penetration of iPhones is substantially higher than for Android phones in the region of the study. We expect that similar results would be obtained with Android phones.

As soon as the SBLE-iOS app is started, and until it is closed, it scans for BLE beacons nearby with the pre-set UUID. When one or more such beacons are found, the app begins collecting data from the sources described below in this section. Beacon scanning and data recording are active even when the app is in the background (e.g., when the user is interacting with another app). In addition, the app is equipped with beacon monitoring capabilities that allow it to detect BLE beacons with the pre-set UUID even when the app was not started (e.g., if the user closed the app, or if the phone was restarted.) In such cases, upon detection of a beacon, a banner is shown on the phone's screen, notifying users that a bus is nearby, and suggesting that they start the app. (Note that data is only recorded when the app is running.) It is important to note that *beacon monitoring*, which is performed if the app was not yet started, is more energy efficient than *beacon ranging*, executed when the app is running. In principle, the beacon monitoring mode could also be used when the app is running, such that, as soon as a beacon is detected, the mode would be switched to ranging, to enable data collection. However, we found that monitoring is substantially less responsive than ranging, with delays that can be as long as a few minutes. For this reason, and considering the data collection nature of our application, the SBLE-iOS app continuously performs ranging while running.

The iOS mobile app collects the following timestamped data:

- BLE beacon's identification (major, minor) and RSSI from the signal received from each beacon with the pre-set UUID, as measured by iOS' Core Location framework. The RSSI is expressed in dBm (decibel-milliwatts), with a minimum value of -99 dBm. RSSI values below -99 dBm are assigned a conventional value of 0.
- Latitude, longitude, velocity (speed and direction) from iOS' Core Location framework. Speed is measured in meters/s, while direction is given as an azimuth angle measured with respect to North.
- Phone's attitude (heading), acceleration, and rotation rate from iOS' Core Motion framework. Note that attitude is measured as a 3-D rotation with respect to a fixed (but unknown) reference system, with the Y axis pointing along the gravity direction.

In addition, the app records all input provided by the user, as explained later. Data from the phone's sensors is sampled every second and stored in a deque (double-ended queue). We chose a low sampling rate to limit energy consumption. Note that, during a sampling period, a participant's phone would read 2 to 3 messages sent by an individual beacon (whose advertisement period, as mentioned earlier, was 350 ms). Every 15 seconds, the content of the deque is transferred through a secure connection to the remote back-end application, where it is stored in the database. Each transfer involves only a few Kbytes of data. In the case of intermittent Internet connection, up to 500 data samples accumulated in the deque can be sent at a time.

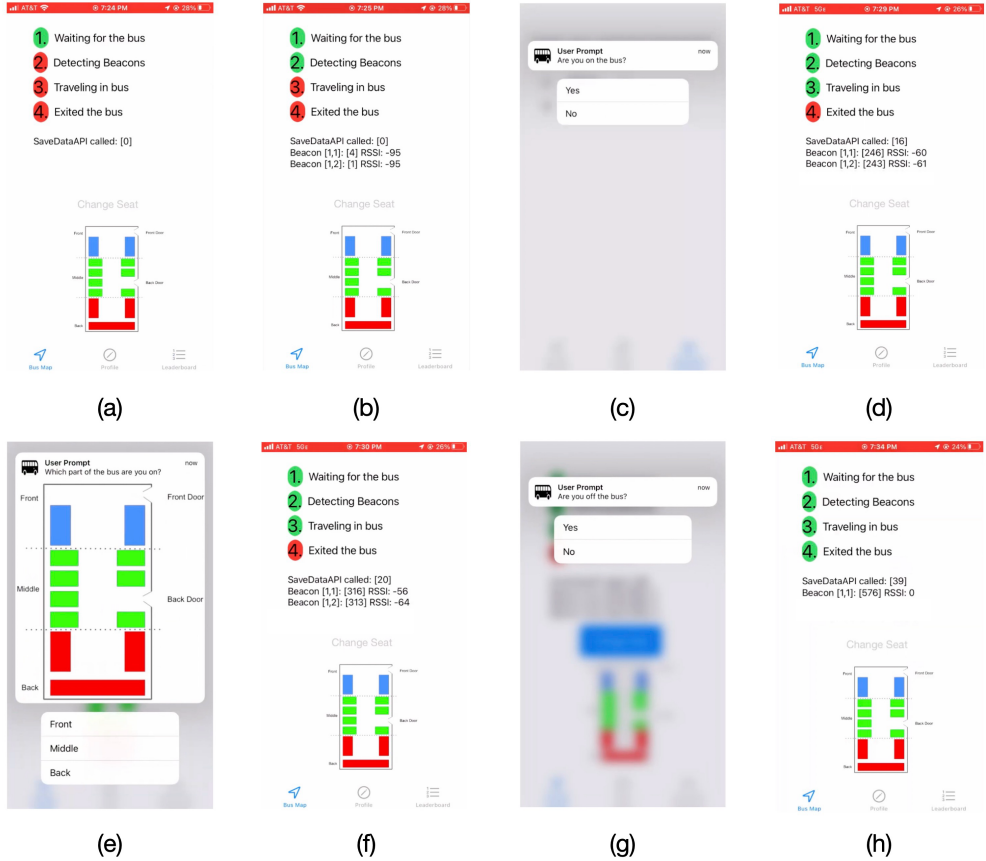

**Figure 2. Screenshots from the SBLE-iOS app displaying the active state and user prompts during a complete sequence.**

At app startup time, users must first login to their account. Referring to Figure 2, the app is in the *Waiting for the bus* state (a), and no data is recorded. As soon as a BLE beacon with the pre-set UUID is discovered, the app switches to the *Detecting Beacons* state (b) and starts recording and transmitting all of the data described above. Note that beacons may be discovered even if the user is not inside a bus, e.g., when another vehicle is passing nearby. If at some point there are no more beacons in range, the app switches back to *Waiting for the bus*. If, however, at least one beacon is within range and the app measures a speed of 5 meters/s or more (indicating that the user is in a moving vehicle), the "Are you in the bus?" notification (c) is produced (via a banner and a vibration), to which the user can reply 'yes' or 'no'. In case the user replies 'no', the same notification is repeated 15 seconds later, if the same conditions are met. If the user replies 'yes', the app switches its state to *Traveling in bus* (c). After 15 seconds, a new notification is produced (e), asking "Which part of the bus are you on?", to which the user can reply with 'Front', 'Middle', or 'Back'. A simple seating map is displayed in the app to help users determine the correct section for their location. If the user moves to another section of the bus, they can select the "Change Seat" button to trigger another seat location notification. When no BLE beacons are in range any more, the app stops collecting data, and produces a further notification "Are you off the bus?" (g). If the user replies 'yes', the app switches to the *Exited the bus* status (h). At this point the app can be safely closed, if the user so desires; otherwise, it switches to the initial *Waiting for the bus* status after a while. All input from the user is timestamped and recorded.

*Distribution and Gamification*
Participation was advertised via email to student mailing lists of two of the largest colleges in the UCSC campus (Baskin School of Engineering and Social Sciences). The app was distributed through TestFlight, an Apple-owned online service that allows over-the-air installation of mobile apps. Interested students were asked to fill out a consent form, after which they were given instructions on how to create an SBLE-iOS account, how to download the TestFlight app and join the program as a tester, and how to install the SBLE-iOS app. When creating an account, each participant was automatically given a nickname, which was used for the leaderboard (to preserve anonymity). Participants who successfully installed the app were offered a $30 gift certificate. In order to incentivize use of the app, we adopted a simple gamification strategy based on a leaderboard. For each participant, the system counted the number of *sessions completed*, where a complete session comprises a sequence of input events from the user, which includes replying to the system prompt for "Are you on the bus?", "Which part of the bus are you on?", and "Are you off the bus?" (causing the system to switch states from *Waiting for the bus* to *Traveling in bus* to *Exited the bus*). The number of sessions completed for each participant (identified by their nickname) in the current month is listed in the leaderboard, which is accessible from the app or from the web site. Each participant can thus see how they fare, in terms of number of completed sessions, as compared to other participants. At the end of each month, one among the top 5 participants in the leaderboard is randomly selected and offered a $150 gift card.

**RESULTS**
The system worked as expected, although one BLE beacon in three shuttle busses (major IDs: 2, 5 and 8) was set to the wrong UUID, meaning that for the 70 trips completed on those busses, our app was only able to record data from the other beacon in the vehicle. 28 students installed the app and completed at least one session (note that data was recorded from all participants, regardless of whether they actually completed a session.) The iPhone model from each participant was also recorded in the data set.

**General Statistics**
Overall, approximately 250,000 data items were recorded, of which about 85% were recorded during 483 complete sessions. Some of the sessions reconstructed from the data were very short, which may be due for example by participants prematurely closing the app. 35 sessions were found to be shorter than 30 seconds, and were removed in our analysis. Figure 3 shows the number of sessions for each participant, along with the average trip length.

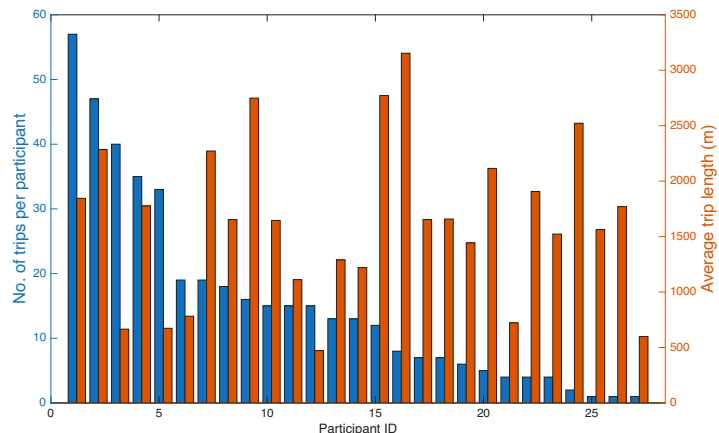

**Figure 3. Number of trips and average trip length per participant.**

Sample paths traversed are shown in Figure 4. As a reminder, a complete session includes data recorded while the system is in the *Detecting beacons* state (beginning when signal from a BLE beacon is detected, which may occur even before a participant enter the bus) and in the following *Traveling in bus* state. The plots in Figure use different colors to indicate the two different system states.

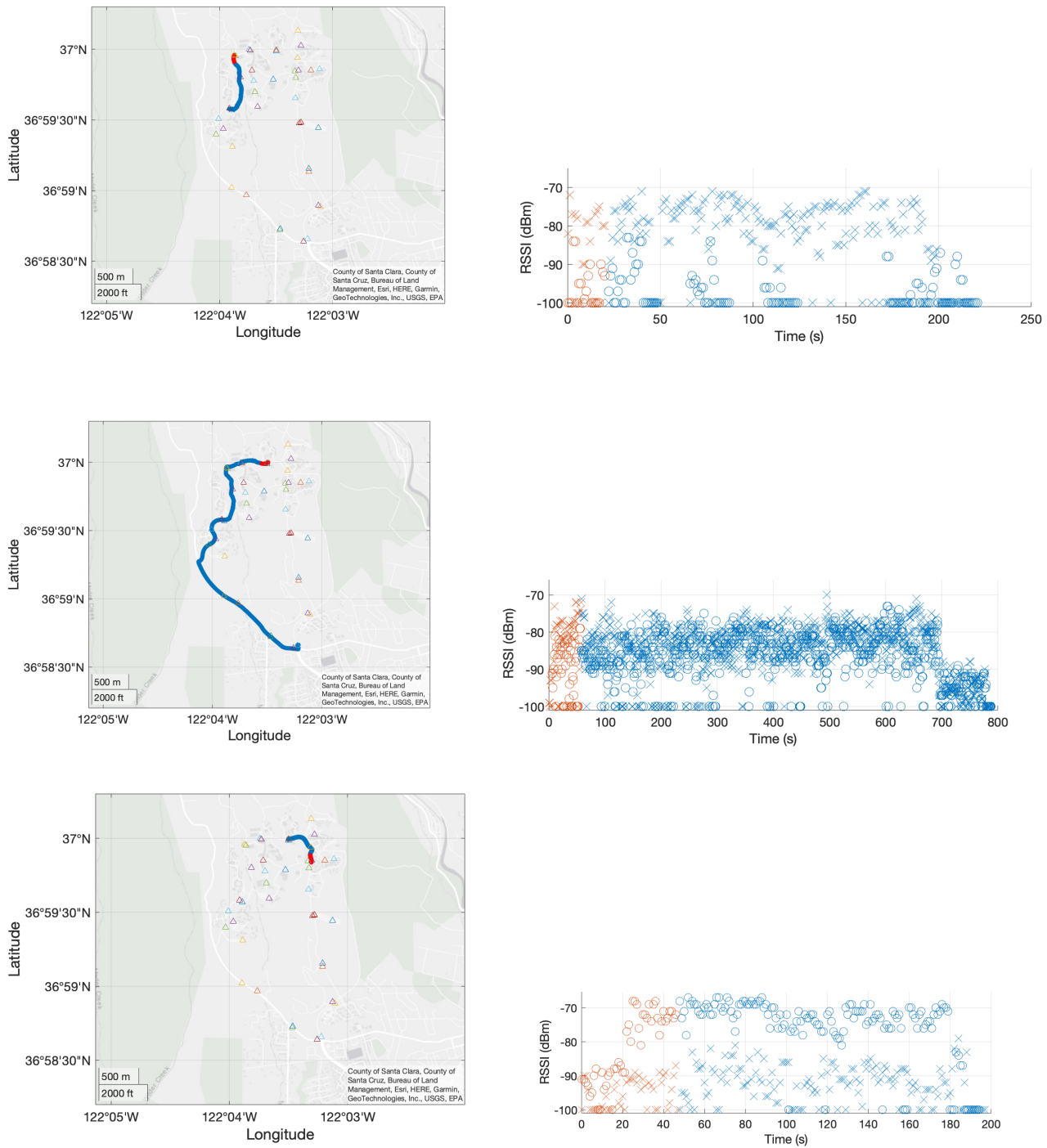

**Figure 4. Representative data collected for participants located in the front section of the bus (top), in the middle section (center), and in the rear section (bottom). For each location, a sample route traversed is shown on the campus map (left), along with the time series of RSSI recorded from the front beacon (crosses) and from the rear beacon (circles). The color of routes and plots represent the state of the system at that time (red: *Detecting beacons*; blue: *Traveling in bus*). The bottom right shape represents the bivariate pdf of RSSI measurements, computed from all recorded data.**

Figure 5 shows the set of all trips, with locations given by the recorded GPS data.

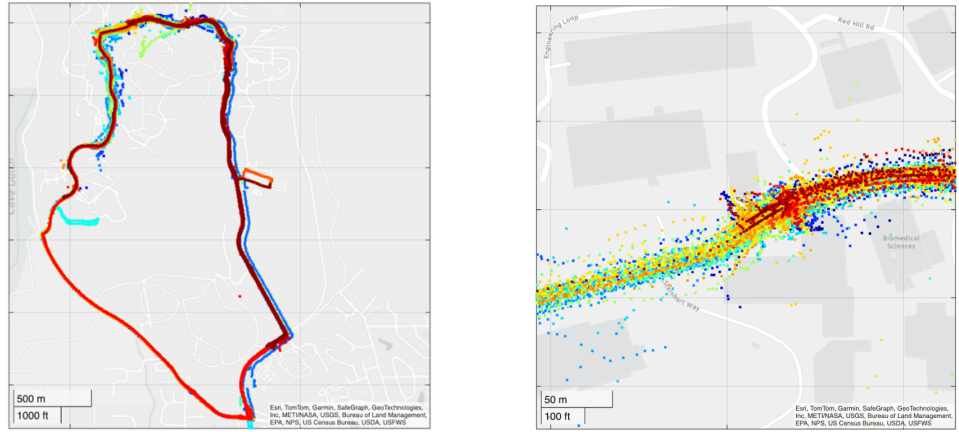

**Figure 5. Left: the set of all trajectories in the recorded sessions. Right: A zoomed-in detail.**

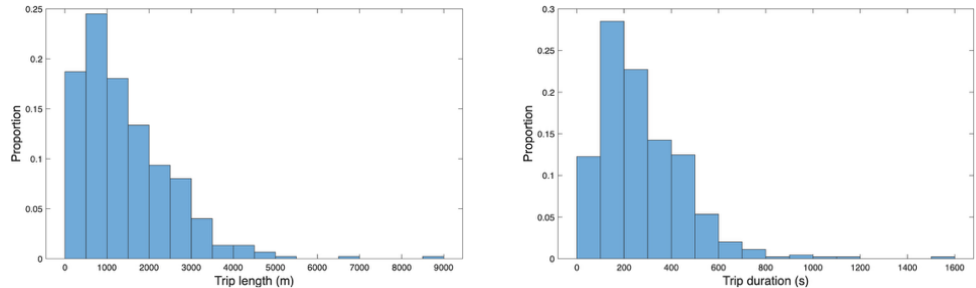

**Figure 6. Histograms of trip lengths (left) and of trip durations (right)**

    The power and advertisement rate of the BLE beacons was chosen as a trade-off between the desire to preserve energy (thus ensuring long battery life) and the need to receive signal from at least one beacon when in the bus. Through prior testing, we determined that the chosen power (-16 dB) was an appropriate compromise. However, sometimes only one beacon was received, or data from one or both beacons had an RSSI value conventionally set to 0 by the iPhone's API, which typically signifies that the received power is lower than -90dBm. (In our analysis, we modified the RSSI for these samples to a default minimum value of -100dBm for further analysis). These situations may occur, for example, if the participant was located at one end of a crowded bus, such that the signal from the beacon at the opposite end could not reach the participant's phone. Figure 7 (left) shows the histogram of RSSI received from either beacon. The bivariate probability density functions (pdf) of the RSSI from the two beacons, computed from all measurements, is shown in Figure 7 (right).

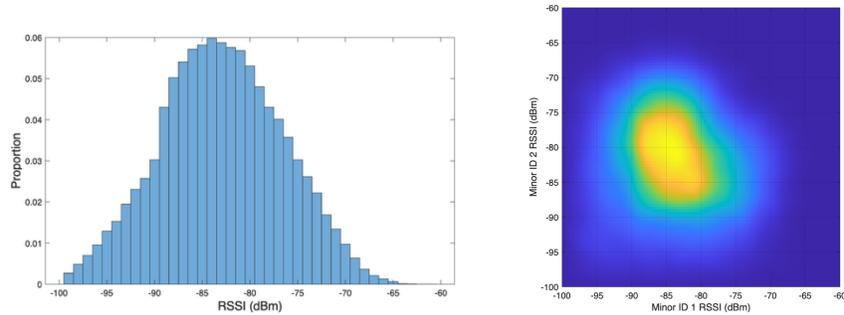

**Figure 7. Left: histogram of received RSSIs. Right: bivariate probability density function of the RSSI from the two beacons.**

One potential source of ambiguity for BIBO systems is when beacons from more than one bus vehicle (hence, with different *major* IDs) are detected at the same time. This occurs when two or more beacon-equipped vehicles are close to the user. In these situations, a BIBO system would need to determine which vehicle (if any) the user is more likely to be in. We found that in 78% of the trips only one major ID was recorded; two majors were recorded in 18% of the trips, and three majors in 3% of the trips.

**Position within the Bus**
As mentioned earlier, the participants' self-reported location in the bus (whether in the front, middle, or back section) was also recorded. This makes it possible to analyze whether the power of the signal received from the two beacons (front and rear) in the same bus correlates with the user's location. Sample RSSI plots from three different trips, with participants in different sections in the bus, are shown in Figure 4. As mentioned earlier, for the sake of visualization, we assigned a value of -100 dBm to samples that were given a conventional value of 0 by the phone. These plots suggest that the user's location does affect the RSSI received from the two beacons. E.g., in Figure 5 (bottom), showing a case with the user in the rear section of the bus, the RSSI from the rear beacon is generally higher than that from the front beacon.

Figure 8 shows the bivariate probability density functions (pdf) of the RSSI from the two beacons for participants sitting in a front, middle, or back seat. Note how the center of mass of these densities is in different locations depending on the seat position in the bus.

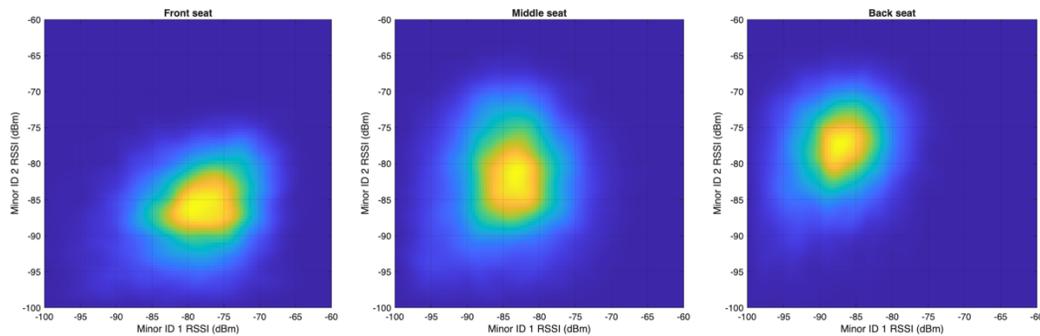

**Figure 8. Bivariate probability density function of the RSSI from the two beacons for participants sitting in a front seat (left), middle seat (center), and back seat (right).**

Another visualization of RSSI distribution is shown in Figure 9, where each dot represents the average RSSI value (from each beacon) for a whole trip. The participant's seat location is shown by the dot's color. This graph shows that the average recorded RSSI from the two beacons can be used to robustly identify whether the user was located in a front or rear seat, but not whether the user was in a middle seat.

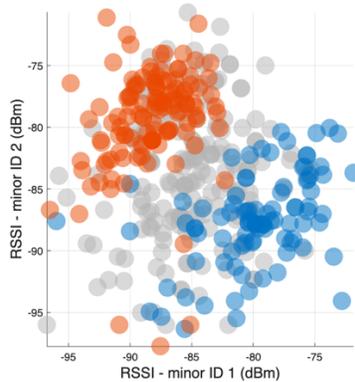

**Figure 9. Scatterplot of average RSSI recorded in each trip for the two beacons when participants were sitting in a front seat (red), middle seat (grey), or back seat (blue).**

**CONCLUSION**
We have presented S-BLE, a data set that can be used for the design and verification of robust BIBO and occupancy systems. S-BLE was collected by multiple participants in a university campus setting. It contains data recorded by the participants' smartphones as they moved around campus, riding shuttle busses that were equipped with BLE beacons. Participants installed an iPhone app (SBLE-iOS) that started recording multiple types of data as soon as a beacon in one of these busses was detected. Data included RSSI from both beacons in each bus, location and odometry, as well as inertial data from the phone's sensors. Participants also attended to notifications generated by the system, which asked to confirm the presence and the approximate location in the vehicle. We presented some statistics of interest related to the RSSI data recorded from the BLE beacons. While not discussed in this contribution, we believe that the inertial data recorded from the participants' smartphones (which includes the phone orientation and acceleration) could also be useful, by providing additional contextual information.

In the following, we highlight some limitations of our approach. It should be clear that this data is only representative of a specific situation (bus vehicles moving in a sparsely urbanized area). Different vehicle types (e.g., train cars) or different environments (e.g., subways) may lead to different signal characteristics. It is also important to note that our data is only *partially labeled*. Data was recorded as soon as a beacon was within range. Only during completed sessions can we assume that participants were inside a bus. As discussed above, the (much larger) amount of off-session data may or may not have been recorded from inside a bus. In fact, while our users were asked to confirm if they were in a bus (because the system detected beacons and measured a speed larger than 5 m/s), they were not asked if they *were not* located inside a bus, e.g. when the system detected a beacon but speed was lower than 5 m/s. This was by design: we decided to limit the rate of system prompts, lest our participants found it too intrusive. Likewise, it is possible (though unlikely) that in some instances no beacon was detected while a participant was riding the bus. Since the app did not provide a way for a user to indicate that they were on the bus without a system prompt, it is impossible, from the recorded data, to ascertain such occurrences. We should remark, though, that even unlabeled data can be very relevant for the design of learning-based BIBO systems (e.g., [14] [13]).

As mentioned earlier, some of our recordings may miss an initial data portion. This is the case, for example, when participants entered a bus while the app was not already running on their phone, which may cause a delayed detection of the beacons. Adding to this delay is the variable amount of time between when a prompt was generated and the user started the app following the prompt. We argue that a similar scenario may occur in a real implementation of a BIBO system, e.g., for fare collection, when the user doesn't have the fare payment app already running on their phone. It is possible, though unlikely, that the delay associated with discovering the beacon while in *beacon monitoring* mode (with the app not yet running) might occasionally have been so long that the whole bus ride went unrecorded.

For what concerns the labeled (in-session) portion of the data, one should consider the possibility that some users might, intentionally or otherwise, have provided wrong information to the app. For example, some participants might have replied incorrectly to the "Which part of the bus are you on?", though this would not give any advantage in terms of leaderboard ranking. While there is no way to identify situations with an incorrect reply, it is possible that someone might have misinterpreted the bus map shown by the app, or, more likely, that they might have changed their location in the bus at some point without bothering to input the new location in the app.

**ACKNOWLEDGMENTS**

This material is based upon work supported by the National Science Foundation under Award No. 2125279.